\documentclass[conference]{IEEEtran}

\usepackage{xcolor}
\usepackage[backend=biber,sorting=none]{biblatex}
\usepackage{graphicx}
\usepackage{siunitx}
\usepackage{dcolumn,tabularx,booktabs}
\usepackage{multirow}
\usepackage{makecell}
\usepackage{subcaption}
\usepackage{circledsteps}
\usepackage{hyperref}
\usepackage{afterpage}
\usepackage{amssymb}
\usepackage{ifthen}

\definecolor{KITgreen}     {RGB}{0,150,130}
\definecolor{KITblue}      {RGB}{70,100,170}
\definecolor{KITblack}     {RGB}{0,0,0}
\definecolor{KITgray}      {gray}{0.3}
\definecolor{KITlightgray} {gray}{0.84}
\definecolor{KITyellow}    {RGB}{252,229,0}
\definecolor{KITorange}    {RGB}{223,155,27}
\definecolor{KITlightgreen}{RGB}{140,182,60}
\definecolor{KITred}       {RGB}{162,34,35}
\definecolor{KITpurple}    {RGB}{163,16,124}
\definecolor{KITbrown}     {RGB}{167,130,46}
\definecolor{KITcyan}      {RGB}{35,161,224}


\makeatletter
\def\ps@IEEEtitlepagestyle{%
  \def\@oddfoot{\mycopyrightnotice}%
  \def\@evenfoot{}%
}
\def\mycopyrightnotice{%
  {\footnotesize
    \parbox{\textwidth}{
      \centering
      \copyright~2026 IEEE. Personal use of this material is permitted. Permission from IEEE must be obtained for all other uses, in any current or future media, including reprinting/republishing this material for advertising or promotional purposes, creating new collective works, for resale or redistribution to servers or lists, or reuse of any copyrighted component of this work in other works by sending a request to \texttt{pubs-permissions@ieee.org}.
    }
  }%
  \gdef\mycopyrightnotice{}
}
\makeatother

\begin{document}

\title{Reconfigurable Computing Challenge: \\ Real-Time Graph Neural Networks for \\ Online Event Selection in Big Science}

\newcommand{\blind}{false}

\DeclareRobustCommand*{\IEEEauthorrefmark}[1]{%
  \raisebox{0pt}[0pt][0pt]{\textsuperscript{\footnotesize #1}}%
}

\ifthenelse{\equal{\blind}{false}}
{
    \author{
    \IEEEauthorblockN{
    Marc Neu\IEEEauthorrefmark{*1},
    Frank Baptist\IEEEauthorrefmark{+},
    Thomas Lobmaier\IEEEauthorrefmark{+},
    Fabio Papagno\IEEEauthorrefmark{*},
    Torben Ferber\IEEEauthorrefmark{+} and
    Jürgen Becker\IEEEauthorrefmark{*}}
    \IEEEauthorblockN{\IEEEauthorrefmark{*}Institute for Information Processing Technology,
    Karlsruhe Institute of Technology, Germany}
    \IEEEauthorblockN{\IEEEauthorrefmark{+}Institute of Experimental Particle Physics,
    Karlsruhe Institute of Technology, Germany}
    \IEEEauthorblockA{
    \IEEEauthorrefmark{1}marc.neu@kit.edu}}
}
{}

\maketitle

\begin{abstract}
Graph neural networks are increasingly adopted in trigger systems for collider experiments, where strict latency and throughput constraints render deployment on embedded platforms challenging.
As detectors move towards higher granularity, the number of inputs per inference increase and FPGA-only solutions face resource bottlenecks.
This work presents an end-to-end demonstrator for the real-time deployment of a dynamic Graph Neural Network for the Belle~II electromagnetic calorimeter hardware trigger on the AMD~Versal~VCK190, leveraging both FPGA fabric and AI~Engine tiles.
We develop a Python-based semi-automated design flow covering operator fusion, partitioning, mapping, spatial parallelization, and kernel-level optimization.
Our design achieves a throughput of 2.94\,million events per second at an end-to-end latency of 7.15\,\textmu{}s.
Compared to the FPGA-only baseline, this represents a 53\,\% throughput improvement while reducing DSP utilization from 99\,\% to 19\,\% at 29\,\% AI~Engine tile utilization.
To validate the deployment, an interactive visualization pipeline enables real-time monitoring of inference results on the physical demonstrator.
\end{abstract}


%
\IEEEpeerreviewmaketitle

\section{Introduction}
\begin{figure}[!t]
    \centering
    \includegraphics[width=\linewidth]{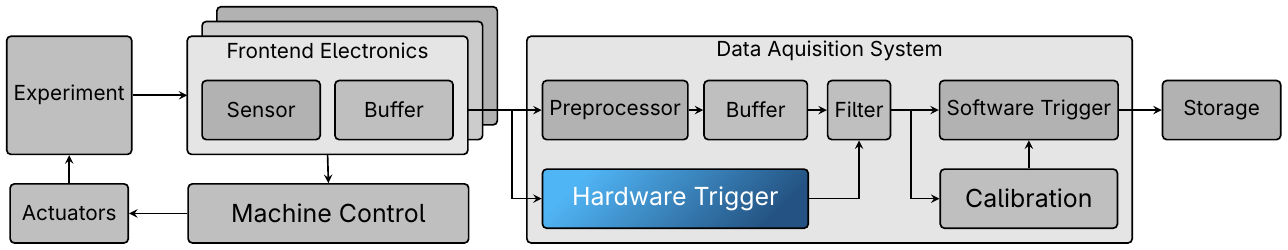}
    \caption{Overview of a typical data acquisition system in large-scale scientific experiments.}
    \label{fig:application}
\end{figure}

Machine learning algorithms are increasingly adopted in first-level triggers in collider experiments~\cite{zipper:2024,Bahr:2024dzg,Liu:2026iup}.
In this context, Graph Neural Networks~(GNNs) require efficient deployment strategies on FPGA-based hardware triggers to satisfy the stringent latency and throughput constraints~\cite{deiana:2022,shlomi:2021,abadal:2022}.
Figure~\ref{fig:application} illustrates a generalized data acquisition system~(DAQ system).
Hardware triggers are typically placed after the frontend electronics in parallel to the preprocessing and readout buffers.
Their purpose is the selection of detector snapshots, so-called events, that are of interest for further offline processing.
Their role as online filter comes with a distinct set of system requirements~\cite{Lai:2025gac,CMS:2000mvk} that make the deployment of state-of-the-art filtering techniques such as GNNs challenging:

(1)~Online filtering decisions must be made within $\SI{10}{\micro\second}$, due to limited buffer capacity.

(2)~Detector snapshots are sampled at rates of up to 40 million events per second, imposing high throughput requirements.

(3)~Trigger decisions must maintain strict in-order processing to ensure correct event-by-event matching with the readout path, requiring hard real-time guarantees for all deployed algorithms.

These requirements are exemplified by the hardware trigger of the Belle~II Experiment~\cite{Belle-II:2010dht} at the SuperKEKB electron-positron collider in Tsukuba, Japan, which serves as the case study in this work.
The experiment studies flavour physics and searches for dark-sector particles at a collision energy of \SI{10.58}{\giga\electronvolt}.
In order to further improve the sensitivity of the experiment, an upgrade of the Electromagnetic Calorimeter~(ECL) is currently discussed.
This upgrade would require significant adaptations to the currently deployed GNN-based trigger system~\cite{haide:2026}:
The current system processes up to 32 nonzero sparse inputs out of 576 per event.
After the upgrade, at least 128 of 8736 nonzero sparse inputs must be considered per event, without a reduction of the overall throughput.

To address this challenge, this work extends the existing FPGA-only GNN baseline to a heterogeneous deployment on the AMD~Versal platform, leveraging the improved performance-to-area ratio of AI Engines (AIEs) in comparison to FPGA fabric~\cite{kuon:2007}.

\section{Related Work}
Two widely adopted frameworks for automated deployment of machine learning algorithms onto FPGAs are FINN~\cite{umuroglu:2017} and hls4ml~\cite{schulte:2025}, the latter of which explicitly targets low-latency applications in experimental particle physics.
Neither supports the deployment of GNNs for real-time applications.
For application-specific deployment, prior work considers CaloClusterNet, a dynamic GNN, for the Belle~II hardware trigger~\cite{haide:2026}, and another study examines the scalability of dynamic GNNs for hardware triggers, providing an open-source High-Level-Synthesis (HLS) library~\cite{neu:2025}.

For deployment onto AMD AI~Engines, the mlir-aie framework provides a low-level programming model~\cite{wang:2025}, but its tight coupling to host-side memory management renders it unsuitable for streaming applications in low-latency hardware triggers.
The aie4ml framework~\cite{danopoulos:2025} extends hls4ml to enable deployment of dense layers onto the AIEs, but supports neither partitioning across FPGA and AIEs nor the deployment of GNNs.

The present work addresses this gap by developing a template library targeting AIEs and providing an end-to-end demonstrator for the deployment of CaloClusterNet onto the heterogeneous AMD Versal System-on-Chip (SoC).
\section{Concept}
\subsection{Deployment}
As a starting point for our work, we consider an existing open-source deployment approach for the CaloClusterNet on FPGA-based hardware triggers~\cite{neu:2025}.
In this reference implementation, all neural network layers are deployed as a dataflow accelerator using HLS.
While this approach offers promising performance, it does not utilize the heterogeneous compute resources available on modern SoC platforms.

To aid the deployment of the CaloClusterNet onto heterogeneous SoCs such as AMD~Versal, we develop a Python-based semi-automated design flow.
Our design flow requires two inputs: a pretrained, quantized neural network model (in our case trained with QKeras~\cite{coelho:2021}), and a set of hardware requirements such as the target throughput and platform.
As output, our design flow emits Vitis~HLS~C\texttt{++} source files for the FPGA partition and Vitis~AIE~C\texttt{++} source files for the AIE partition.

Our deployment flow is separated into a series of stages: operator fusion, partitioning, mapping, spatial parallelization, and kernel-level optimizations.
Internally, we represent the neural network as a dataflow graph, where nodes correspond to operators (individual layers) and edges describe data dependencies.
Each stage transforms this graph until it is lowered to platform-specific source files.
In the following, we describe each step.

\begin{figure}[h]
    \centering
    \includegraphics[width=0.8\linewidth]{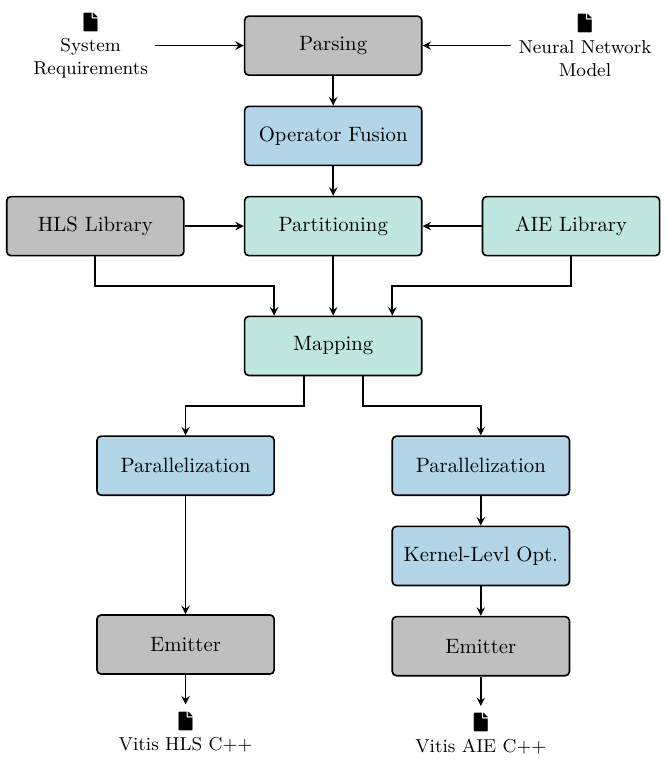}
    \caption{Overview of our deployment flow. Key transformation stages are green. Optimization stages are blue.  All others stages are gray.}
    \label{fig:deployment}
\end{figure}

\textit{Operator Fusion.}
First, Linear layers and their subsequent ReLU activations are fused into a single Dense operator.
Second, parallel Dense operators with the same predecessor are merged into a single Dense operator whose output features are the concatenation of the individual operators' outputs.
These optimizations simplify the dataflow graph topology by removing the multicast on the preceding operator.
The simplified topology is particularly critical, as no more than eight AIE memory buffers are connected to each tile on the first generation of AMD~Versal.
Introducing even a single multicast requires four memory buffers due to double buffering, leaving insufficient buffers for the remaining operators during tile placement.

\begin{figure*}[p]
    \centering
    \includegraphics[width=\linewidth]{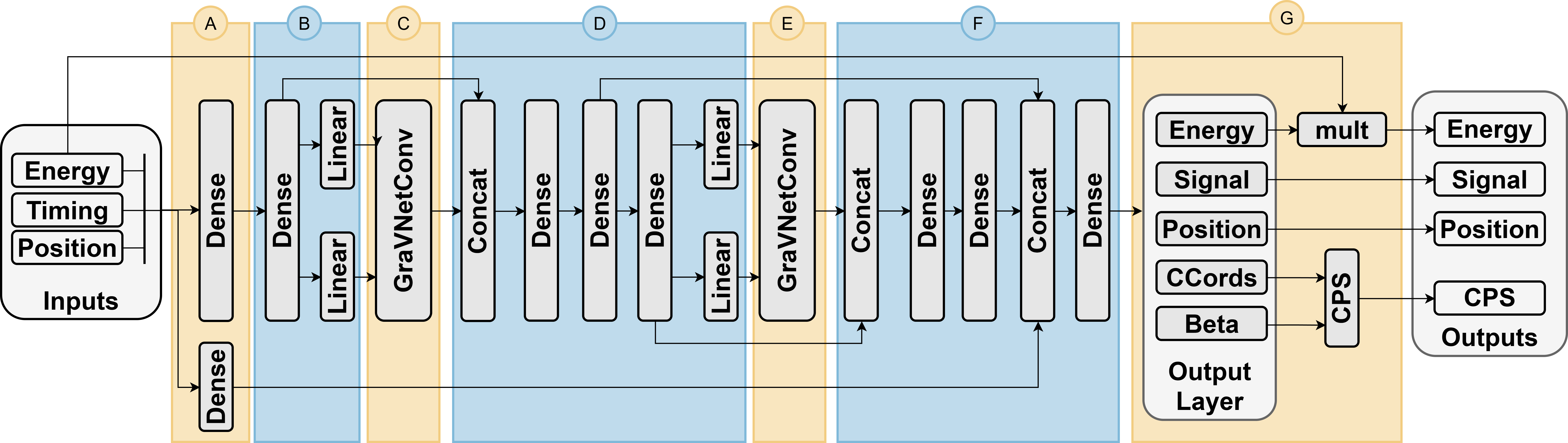}
    \caption{The CaloClusterNet after partitioning onto the AMD~Versal architecture. Partitions implemented on the FPGA are shown in yellow. Partitions implemented on the AI~Engine fabric are shown in red.}
    \label{fig:caloclusternet}
\end{figure*}

\textit{Partitioning.}
Next, we assign each operator to one of two target platforms: FPGA logic or AIE tiles.
We employ a greedy scheme that prioritizes the AIE platform, as AIE tiles offer better performance-per-area.
All operators with regular, statically scheduled data access patterns are assigned to the AIE platform, including layers such as Linear, Dense, ReLU, or Concat, but excluding GNN-specific layers which require data-dependent memory access.
All AIE kernels run fully decoupled, without global memory access, avoiding memory stalls at runtime.
The remaining operators, including input and output layers interfacing with DDR~RAM, are assigned to FPGA logic.
As only operators with regular access patterns are eligible for AIE assignment, the space of valid configurations is small and the greedy scheme is sufficient without exhaustive search.
The resulting partitioning is shown in Figure~\ref{fig:caloclusternet}: seven segments are derived, of which four are implemented on FPGA and three on AIEs.

\textit{Mapping.}
After partitioning, each operator is mapped to a corresponding architecture template through pattern matching.
For FPGA operators, we use the HLS templates from~\cite{neu:2025}, which support the irregular, data-dependent access patterns required by graph convolutions (\textit{GravNetConv}) and the Condensation Point Selection (\textit{CPS}) algorithm.
For AIE operators, we develop an open-source, header-only C\texttt{++} kernel library for the AMD chess compiler, providing reusable templates for Dense, Linear, Concat, and ReLU operators.
During mapping, the dataflow graph is also legalized: for each edge, if the output layout of the source operator does not match the expected input layout of the sink operator, a Retile kernel is inserted to reshape the intermediate tensor.
Each AIE template encapsulates a single operator as a self-contained kernel.

\textit{Spatial Parallelization.}
After mapping, each AIE partition is optimized through spatial parallelization.
GNN layers that do not perform feature aggregation across neighbors are fully spatially separable, meaning they can process independent spatial regions in parallel.
The operator fusion step further improves separability by reducing each AIE partition to a linear chain of operators, eliminating cross-partition data dependencies.
For each partition, we select a spatial parallelization factor $P \in \{2^n \mid n \in \mathbb{N}_0\}$, which replicates the operator chain $P$ times.
On the AIE, resource utilization scales linearly with $P$; on the FPGA, growth is superlinear due to additional routing and buffering overhead.
We apply an exhaustive search to find the smallest $P$ that satisfies the target throughput, minimizing resource utilization.

\textit{Kernel-Level Optimizations.}
Finally, we optimize individual AIE kernels to meet the strict timing constraints of our application.
To sustain a throughput of 1\,MHz, each kernel must complete within 1\,\textmu s.
At this scale, loop pipelining overhead imposed by the \texttt{chess} compiler becomes significant relative to the total kernel runtime.
The AMD Vitis AIE library~\cite{vitis_libraries:2024} amortizes pipelining setup cost over many iterations, which is inefficient for the small matrix dimensions used here.
In our kernel library, we replace loop pipelining with loop flattening via \texttt{chess\_flatten\_loop} annotations, trading larger program memory for better performance.
The impact of this optimization is evaluated in Section~\ref{sec:evaluation}.

To evaluate the impact of each optimization stage, we define three successive design points.
Design~\Circled{1} serves as a baseline, implementing the partitioned network without further optimizations.
Design~\Circled{2} applies operator fusion and spatial parallelization with $P_{\mathrm{fpga}}=2$ and $P_{\mathrm{aie}}=4$.
Design~\Circled{3} additionally applies the kernel-level optimizations described above.

\begin{figure}[h]
    \centering
    \includegraphics[width=\linewidth]{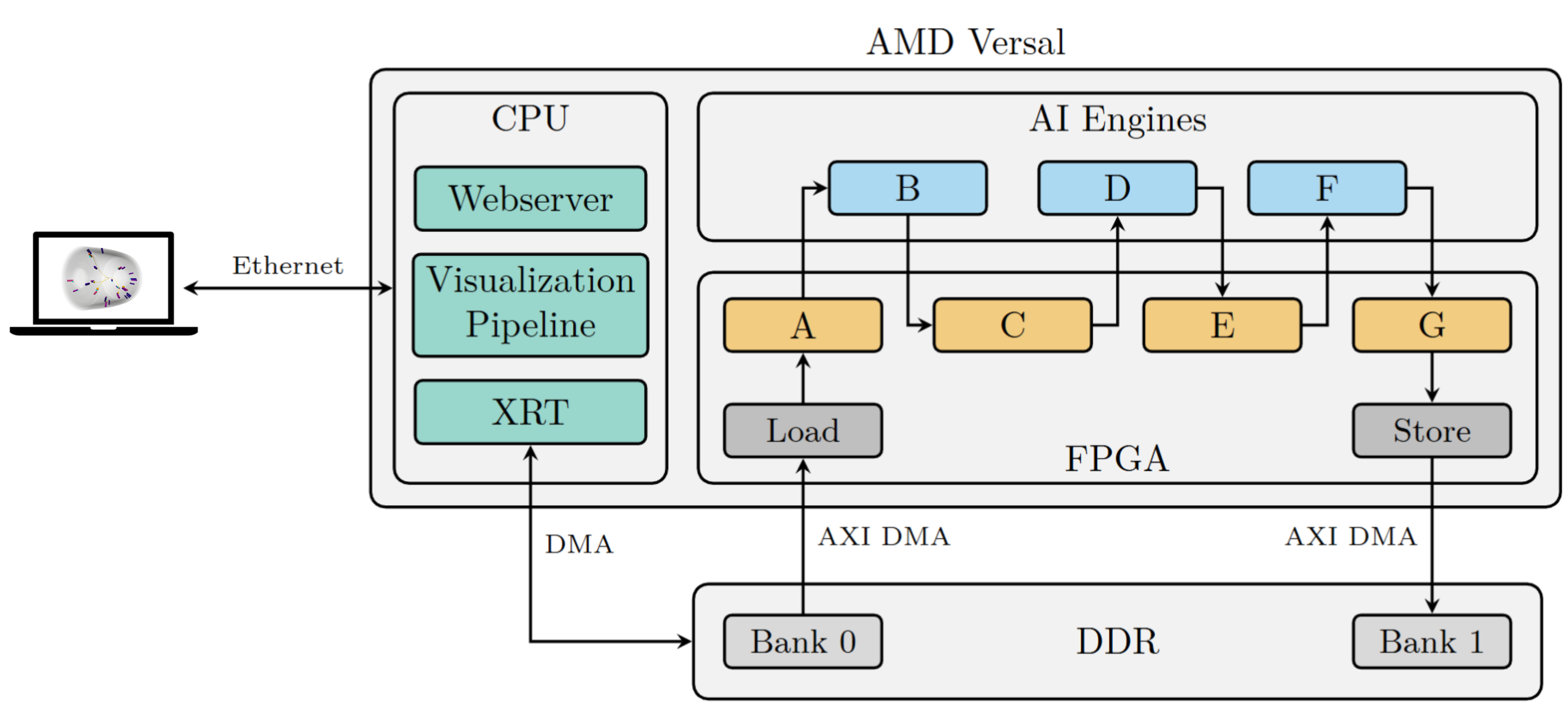}
    \caption{System architecture of our demonstrator on the AMD~Versal~VCK190.}
    \label{fig:architecture}
\end{figure}

\subsection{System Architecture}

We implement an end-to-end demonstrator on the AMD~Versal~VCK190, shown in Figure~\ref{fig:architecture}.
The system comprises three partitions: the Arm~Cortex~A72 CPU, the FPGA, and the AIE array.
Input data is read from DDR Bank~0 by the Load kernel on the FPGA, which feeds the first compute kernel of the accelerator pipeline.
The seven compute kernels (A~--~G) alternate between FPGA and AIE partitions, forming a dataflow pipeline that processes inference requests without CPU intervention.
Results are written back to DDR Bank~1 by the Store kernel.
On the CPU, the Xilinx Runtime (XRT) manages kernel orchestration and DMA transfers.
A visualization pipeline postprocesses the inference results and serves them to an external client via an integrated webserver over Ethernet, rendering a three-dimensional event display of the Belle~II ECL detector with interactive selection among common physics-case datasets at varying background levels.
\section{Performance Analysis}
\label{sec:evaluation}
For evaluation, we implement our end-to-end demonstrator on the AMD~Versal~VCK190 evaluation board using AMD~Vitis~2024.2 and AMD~Vivado~2024.2.
Profiling monitors are inserted on the Programmable Logic partition for all kernels to record execution timelines.
AIE kernel execution times are derived indirectly from the corresponding PLIO interface timestamps.
All kernels on Programmable Logic operate at $f_{\mathrm{fpga}} = 250\,\mathrm{MHz}$, and kernels on AIE at $f_{\mathrm{aie}}=1.25\,\mathrm{GHz}$.
To reach timing closure, the \texttt{\detokenize{Flow_PerfOptimized_High}} synthesis strategy and \texttt{\detokenize{Performance_ExplorePostRoutePhysOpt}} implementation strategy are used in AMD~Vivado.

Functional correctness is validated in software using QKeras, in hardware emulation, and on hardware, with bit-accurate agreement across all three.

We compare our implementation against the FPGA-only baseline from~\cite{neu:2025} on the AMD~Versal~VCK190; a GPU reference on an NVIDIA~L40S using TensorRT, is included in the figures as baseline.
All implementations support up to 128 inputs per inference; fewer inputs are supported through zero-padding.
The FPGA-only version uses 8-bit quantization for all layers.
The FPGA \& AIE versions use 16-bit precision for partitions A and G to preserve inference quality at the system boundaries, and 8-bit for all remaining partitions.

\begin{figure*}[p]
    \vspace*{0pt}
    \centering
    \begin{subfigure}{\columnwidth}
        \includegraphics[width=\columnwidth]{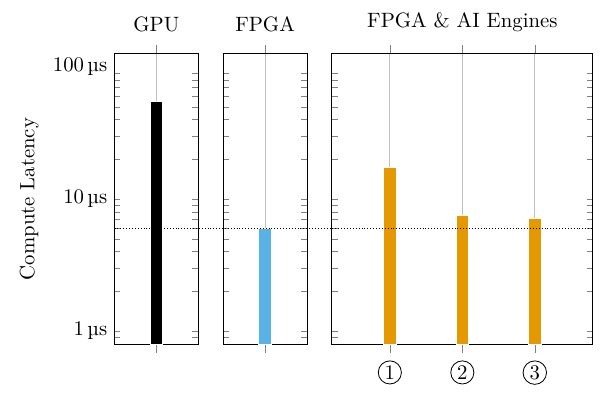}
        \caption{End-to-end latency.}
        \label{fig:latency}
    \end{subfigure}
    \hfill
    \begin{subfigure}{\columnwidth}
        \includegraphics[width=\columnwidth]{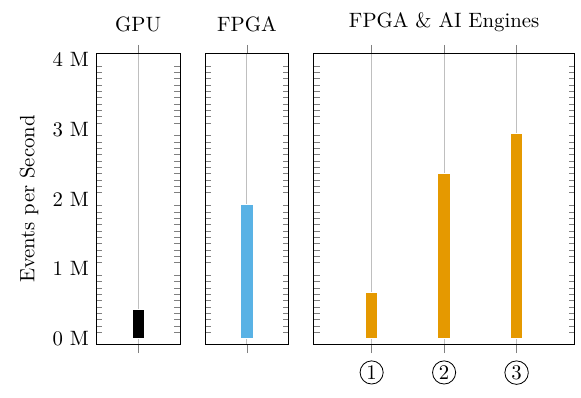}
        \caption{End-to-end throughput.}
        \label{fig:throughput}
    \end{subfigure}
    \caption{Performance evaluation for various versions of the hardware accelerator.}
    \label{fig:performance}
\end{figure*}

Performance results are depicted in Figure~\ref{fig:latency} and Figure~\ref{fig:throughput}.
Our initial design~\Circled{1} performs worse than the FPGA baseline in both latency and throughput, due to overhead introduced by the heterogeneous partitioning.
After optimization, design~\Circled{2} achieves a throughput of 2.36\,million events per second at a latency of \SI{7.47}{\micro\second}.
The best performance is achieved by design~\Circled{3}, with a throughput of 2.94\,million events per second and a latency of \SI{7.15}{\micro\second}.
Relative to the FPGA-only baseline from~\cite{neu:2025}, design~\Circled{3} achieves a throughput improvement of 53\,\% at a latency overhead of 18\,\%.
Designs~\Circled{2} and~\Circled{3} share identical tile allocation.
The improvement in design~\Circled{3} results solely from kernel-level optimization which reduces execution time without altering resource counts.

System resource utilization is shown in Table~\ref{tab:resource_utilization}.
The FPGA baseline uses 65~\% of available lookup tables and 99~\% of digital signal processors.
Design~\Circled{3} requires 53~\% of lookup tables and 19~\% of digital signal processors, at a cost of 29~\% of available AI~Engine tiles.

\section{Conclusion}
In this work, we presented an interactive end-to-end GNN hardware accelerator on the AMD~Versal~VCK190 for the Belle~II ECL hardware trigger. 
Our design achieves a 53\% throughput improvement at 18\% latency overhead over the FPGA-only baseline. 
All source code and device images are publicly available, making our demonstrator a practical reference design for real-time GNN deployment on AMD~Versal SoCs.

\clearpage

\begin{table*}[p]
  \centering
  \caption{System resource utilization for various design implementations on the AMD~Versal~VCK190 Evaluation Board.}
  \label{tab:resource_utilization}
  \small
  \begin{tabular*}{\textwidth}{
    @{\extracolsep{\fill}}
    lll
    S[table-format=6.0] S[table-format=2.0]
    S[table-format=6.0] S[table-format=2.0]
    S[table-format=4.0] S[table-format=2.0]
    S[table-format=4.0] S[table-format=2.0]
    S[table-format=3.0] S[table-format=2.0]
    S[table-format=3.0] S[table-format=2.0]
    S[table-format=3.0] S[table-format=2.0]
  }
    \toprule
    & & &
    \multicolumn{8}{c}{FPGA} &
    \multicolumn{6}{c}{AI Engines} \\
    \cmidrule(lr){4-11} \cmidrule(lr){12-17}
    & & &
    \multicolumn{2}{c}{FF} &
    \multicolumn{2}{c}{LUT} &
    \multicolumn{2}{c}{DSP} &
    \multicolumn{2}{c}{BRAM} &
    \multicolumn{2}{c}{Tiles} &
    \multicolumn{2}{c}{Compute} &
    \multicolumn{2}{c}{Memory} \\
    \cmidrule(lr){4-5} \cmidrule(lr){6-7} \cmidrule(lr){8-9} \cmidrule(lr){10-11}
    \cmidrule(lr){12-13} \cmidrule(lr){14-15} \cmidrule(lr){16-17}
    Target & Precision & Iter. &
    {abs.} & {\%} & {abs.} & {\%} & {abs.} & {\%} & {abs.} & {\%} &
    {abs.} & {\%} & {abs.} & {\%} & {abs.} & {\%} \\
    \midrule
    FPGA        & 8\,bit & {--}           & 625808 & 36 & 573300 & 65 & 1951 & 99 & 170 & 19 & {--} & {--} & {--} & {--} & {--} & {--} \\
    \midrule
    FPGA \& AIE & mixed  & \Circled{1}    & 384395 & 22 & 287284 & 33 & 186  & 9  & 98  & 10 & 39  & 10  & 25 & 6  & 39  & 10 \\
    FPGA \& AIE & mixed  & \Circled{2}    & 484210 & 28 & 462972 & 53 & 372  & 19 & 292 & 33 & 117 & 29 & 84 & 21 & 117 & 29 \\
    FPGA \& AIE & mixed  & \Circled{3}    & 484210 & 28 & 462972 & 53 & 372  & 19 & 292 & 33 & 117 & 29 & 84 & 21 & 117 & 29 \\
    \bottomrule
  \end{tabular*}
\end{table*}%
\vfill
\clearpage

\section*{Code Availability Statement}
The complete source code, included our performance measurements, device image, and visualization pipeline is available on GitHub: \href{https://github.com/marcneu/vck190-caloclusternet-demo}{https://github.com/marcneu/vck190-caloclusternet-demo}.

\section*{Acknowledgment}
This work was supported through the DEEP consortium (05D25VK1) funded by the German Federal Ministry of Research, Technology and Space (BMFTR) in the ErUM-Data action plan.
    
\printbibliography

\end{document}